\documentstyle[aps,pra,twocolumn,floats,epsfig]{revtex}

\newcommand{\eins}{\mbox{$1 \hspace{-1.0mm}  {\bf l}$}}
\newcommand{\be}{\begin{equation}}
\newcommand{\ee}{\end{equation}}
\newcommand{\bea}{\begin{eqnarray}}
\newcommand{\eea}{\end{eqnarray}}
\def\bma{\begin{mathletters}}
\def\ema{\end{mathletters}}
\newcommand{\half}{\mbox{$\textstyle \frac{1}{2}$}}

\newcommand{\shalf}{\mbox{$\textstyle \frac{1}{\sqrt{2}}$}}
\newcommand{\ket}[1]{ | \, #1  \rangle}
\newcommand{\bra}[1]{ \langle #1 \,  |}
\newcommand{\braket}[2]{\left< #1 \right| #2 \rangle}
\newcommand{\proj}[1]{\ket{#1}\bra{#1}}

\begin{document}
\draft

\title{Detection of entanglement with  few local measurements}
\author{O. G\"uhne$^1$, P. Hyllus$^1$, D. Bru\ss$^1$, A.~Ekert$^2$,
M. Lewenstein$^1$, 
C.~Macchiavello$^3$, and A. Sanpera$^1$}
\address{
$^1$Institut f\"ur Theoretische Physik, Universit\"at Hannover, 30167
Hannover, Germany\\
$^2$Dep. of Appl. Math. and Theoret. Phys., University of
Cambridge,
Wilberforce Road, Cambridge CB3 0WA, UK
\\
$^3$Dipartimento di Fisica ``A. Volta" and INFM-Unit\'a 
di Pavia,
Via Bassi 6, 27100 Pavia, Italy}
\date{Received \today}

\maketitle
\begin{abstract}
We introduce a general method for the experimental
detection of entanglement by performing only  few local 
measurements, assuming  some prior knowledge of the density
matrix. The idea is based on the minimal decomposition of
witness operators into a pseudo-mixture of local operators.  
We discuss an experimentally relevant case of two qubits, 
and show an example how bound entanglement can be 
detected with few local measurements. 
\end{abstract}


\narrowtext

A central aim in the physics of quantum information is to
create and detect entanglement -- the resource that allows to realize
various quantum protocols. Recently, much progress has been achieved 
experimentally in creating entangled states \cite{experiments}. 
In every real experiment noise and imperfections are present so that
the generated states, although intended to be entangled, 
may in fact be separable. Therefore, it is important to find
efficient experimental methods to test whether a given imperfect state 
$\rho$ is indeed entangled. 

Obviously, the ultimate goal of entanglement detection is to  characterize
entanglement quantitatively, and identify regions in 
the parameter space which
allow to maximize entanglement  for a particular quantum
information processing task. The first step towards this ambitious goal
is to detect whether a given state is entangled or not.

The question of direct detection of quantum entanglement has been recently
addressed in Refs. \cite{susana,paar,pawel}. In \cite{paar,pawel} 
the authors study the case of mixed states and find efficient ways 
to estimate the entanglement of an unknown state. 
Their method is based on structural approximations of some 
linear maps followed by a spectrum estimation.
Although experimentally viable the method is not very easy to implement and
it requires further modifications in order to be performed by local
measurements \cite{private}. Here, we approach the
same problem from a different perspective. We use special observables, the
so-called witness operators \cite{terwit,opti} and their optimal
decomposition into a sum of {\em local} projectors. 
Note that in this way we answer an open question posed recently
in \cite{filip}, where {\em non-local} measurements of entanglement
witnesses were studied.

The construction of a witness 
for a given arbitrary state is, in general, a formidable task. It can, 
however, be accomplished  in typical experimental  situations
where one has some {\it a priori} information about 
the density matrix. This is always the case when the experiment is 
{\it aimed} at producing a certain state, rather than checking 
properties of an {\it a priori} unknown state. 
We  discuss two experimentally relevant situations in this paper, 
namely the generation of a definite pure entangled state of two 
parties, and the generation of a specific bound entangled edge state. 
In both cases our method can be applied in arbitrary dimensions.

Having constructed a witness, its measurement can be performed {\it locally}, 
since every observable can be decomposed in terms of a
product basis in the operator space. 
Here we  propose two ways of optimizing 
such local measurements. The first one consists in looking for  
the optimal number of local projectors (ONP).  
The second one consists in searching  for the optimal 
number of  settings of detecting devices (ONS). 
By a setting of the device of a single observer 
we understand here the choice of the local orthonormal basis in the 
corresponding Hilbert space. The device measures then
simultaneously projections onto the vectors belonging to the  basis;  the set
of these projectors forms a complete set of commuting
observables \cite{stern}. A setting of the devices for a pair of 
observers corresponds then to a correlated choice of individual
settings.

Both optimization methods are formulated as the problem 
of decomposing a given operator into a sum of projectors on product 
states with an optimal number of terms. No general solutions for this 
kind of problems are known so far. 

Before describing the details of our method, let us briefly discuss 
other methods of entanglement detection with local measurements. 
For systems of two qubits or of  one qubit and one qutrit, a necessary 
and sufficient criterion for entanglement, namely the non-positivity 
of the  partial transpose \cite{peres}, is known. Thus, using 
to\-mo\-gra\-phy 
of the state $\rho$, which can be achieved with  local measurements 
\cite{tomo},  one can fully determine $\rho$, calculate the partial 
transpose and check its positivity. However, for two qubits this approach 
requires 9 different settings of the measuring devices in order to 
determine 15 parameters describing the state in general. In our example 
below, only 3 settings suffice to detect entanglement if we have 
certain knowledge about $\rho$ and we optimize the local decomposition 
of the corresponding witness.  Our approach has similar advantages 
with respect to the detection of {\it entanglement visibility}\cite{evis}, 
which in principle requires a continuous family of devices' settings. 

Another way of detecting entanglement 
by local measurements consists in  a test of a 
generalized Bell inequality\cite{bell,terbell}, with which our 
approach has a formal similarity. Nevertheless there exist many entangled 
states which do not violate any known Bell inequality \cite{wolf1}. 
However, for any given state one can {\em always} find a witness operator 
and its local decomposition, such that the entanglement of this state 
can be  detected locally. For the situations, 
in which 
a previous knowledge 
of $\rho$ allows to construct  a suitable witness, 
our method is more powerful than a test of a Bell inequality. 

We introduce and illustrate our method in the scenario of 
creating and measuring the entanglement of two qubits, 
and then discuss shortly the generalisation to higher-dimensional states, 
including bound entangled states. 

Let us consider an experiment that 
produces the following convex combination of a desired pure entangled 
state $\proj{\psi}$ and a mixed state $\sigma$ representing some noise,
\begin{equation}
\varrho = p\proj{\psi}+(1-p)\sigma\ , \ \ \ 0\leq p \leq 1\ ,
\label{rho}
\end{equation}
where $\ket{\psi}$ 
can be written 
in the Schmidt decomposition as $\ket{\psi}=a \ket{01}+b\ket{10}$ with $a,b>0$ 
and $a^2+b^2=1$.  The noise $\sigma $ is assumed to be within a ball of 
radius $d$ around the totally mixed state, i.e.
$||\sigma - \eins/4 ||\leq d$. Here
$\Vert A \Vert := \sqrt{Tr(A^{\dagger}A)}$ is the Hilbert-Schmidt norm for 
operators $A$ on the Hilbert space. 
One neither knows the probability 
$p$, nor the exact shape of $\sigma$. The task is to determine whether 
$\varrho$ is entangled or not. 

Let us briefly summarise the well-known concept of 
entanglement witnesses \cite{terwit,opti}: a density matrix $\varrho$ is 
entangled iff there exists a Hermitian operator $W$ such that 
$Tr(W\varrho)<0$, but for all separable states $Tr(W\varrho_{sep})\ge 0$ 
holds. In this sense $W$ ``detects" the entanglement of $\varrho$. Note 
that $W$ has at least one negative eigenvalue. Methods to construct 
entanglement witnesses have been presented in Refs. \cite{terwit,opti}. 
For states with a non-positive partial transpose there is a simple and 
straightforward construction of $W$: 
let $\ket{e_-}$ be the eigenvector 
of $\varrho^{T_A}$ that corresponds to its minimal (negative) eigenvalue, 
namely $\varrho^{T_A}\ket{e_-}=\lambda_{min}\ket{e_-}$, with 
$\lambda_{min}<0$. 
Here $T_{A}$ refers to partial transposition with respect to the 
first subsystem.
Thus, $W=(\proj{e_-})^{T_A}$ detects the entanglement 
of $\varrho$, as 
$Tr((\proj{e_-})^{T_A}\varrho)=Tr(\proj{e_-}\varrho^{T_A})=\lambda_{min}<0$.
This witness is tangent to the set of separable states and already
optimal \cite{opti}, {\em i.e.} there is no witness that 
detects other states in addition to the ones detected by $W$. 

For $\varrho$ given in Eq. (\ref{rho}) 
and the case $d=0$ one finds $\lambda_{min}=(1-p)/4-pab,$ and
$\ket{e_-}=\shalf(\ket{00}-\ket{11})$, {\it i.e.} the witness is given by
\begin{equation}
W = \frac{1}{2} 
\left( 
\begin{array}{cccc}
1 & 0 & 0 & 0\\
0 & 0 &-1 & 0\\
0 &-1 & 0 & 0\\
0 & 0 & 0 & 1
\end{array}
\right). 
\label{witness}
\end{equation}
Note that this witness  neither depends on $p$, nor on 
$a$. Hence as long as the experimental apparatus produces
{\em any} superposition of $\ket{01}$ and $\ket{10}$ plus
white noise, $W$ will be a suitable operator. For $d\neq 0$ 
it still provides the possibility of entanglement detection. 

Let us point out that this witness is also suitable for other 
physical scenarios. For example, consider the case where the noise 
mechanisms in the experimental setup are characterised by memory effects, 
or the case where the entangled state $\ket{\psi}$, (defined via Eq. 
(\ref{rho})), is  generated perfectly and then sent through a 
transmission channel with correlated noise.  If the noise mechanisms 
acting on the  state can be described as a  depolarising channel with some 
correlations of strength $\mu$ \cite{memory}, then the resulting state will 
be of the form:
\begin{eqnarray} 
\varrho&=&\left\{\eins\otimes \eins+\eta(a^2-b^2)[\sigma_z\otimes \eins-
\eins\otimes \sigma_z]\right.\label{memory}
\\
&+&[\mu+(1-\mu)\eta^2][-\sigma_z\otimes\sigma_z\;\nonumber\\
&+&\left. 2ab(\sigma_x\otimes\sigma_x+
\sigma_y\otimes\sigma_y)]\right\}/4\;.\nonumber
\end{eqnarray} 
Here $\eta$ and $\mu$ describe the depolarisation and
the degree of memory introduced by the noise process. 
This family of states is now characterized by three independent parameters:
$a$, $\eta$ and $\mu$.  Remarkably, $W$ turns out to detect the entanglement 
of this whole family of states. This is proven by calculating the 
range of the family's three
 parameters where $\varrho^{T_A}$ has a negative eigenvalue, and showing 
that for this range $Tr(W\varrho)<0$ holds.

We look now for a decomposition of the witness into a sum of projectors 
onto product vectors, {\em i.e.}
\begin{equation}
W=\sum_{i}c_{i}\proj{a_{i},b_{i}} 
 = \sum_{i}c_{i}\proj{a_{i}}\otimes\proj{b_{i}}\ ,
\label{deco}
\end{equation}
where the coefficients $c_{i}$ are real and fulfill $\sum_ic_{i}=1$.  
Note that at least one coefficient has to be negative -- this 
characterises a so-called {\it pseudo-mixture}. Any bipartite
Hermitian operator can be decomposed in projectors onto product states,
like in Eq. (\ref{deco}), in many different ways. However, we are interested 
in finding the {\em optimal} decompositions in the two ways described above.
Optimal pseudo-mixtures 
in the sense of minimizing the number of non-vanishing $c_i$ represent an 
ONP. They have been studied for two qubits in \cite{sanpera}
where it was shown  that any general vector 
$\ket{\phi}=\alpha \ket{00}+\beta \ket{11}$ with 
$\alpha, \beta$ real and  different from zero, and
$\alpha^2+\beta^2=1$, can be decomposed minimally with 5 terms:

\begin{eqnarray}
(\proj{\phi})^{T_A}&=&\frac{(\alpha+\beta)^2}{3}\sum_{i=1}^3 \proj{ f_i  f_i} 
\nonumber
\\
&& - \alpha \beta (\proj{01}+\proj{10})
\end{eqnarray}
where we have used the definitions
\begin{eqnarray}
\ket{f_1}& = & e^{-i\frac{\pi}{3}} \cos\theta \ket{0} 
                    + e^{i\frac{\pi}{3}} \sin\theta\ket{1} 
= \ket{{f_2}^{*}}
\nonumber \\
\ket{f_3}& = & \cos\theta \ket{0}+\sin\theta\ket{1} 
  =\ket{f_1}+\ket{f_2}
\nonumber \\
\cos\theta&=&\sqrt{\alpha/(\alpha+\beta)}, 
\hspace{0.3cm} \sin\theta=\sqrt{\beta/(\alpha+\beta)}.
\end{eqnarray}
Here $*$ denotes complex conjugation.
The case  $\alpha=\shalf=-\beta$ corresponds exactly to 
the decomposition of $W$ in Eq. (\ref{witness}) that we are looking for. 
Such a decompostion into 5 terms requires four different settings of the
measuring device (in the sense described above and in the footnote
\cite{stern}).
It is therefore optimal in the number of projectors (ONP),
but {\em not} optimal with respect to the number of correlated 
devices' settings  (ONS). Indeed, the witness $W$  considered 
in Eq. (\ref{witness}) can be optimally implemented by using only 
three settings. 
This can be shown as follows:
defining the eigenstates of the Pauli 
matrices as $\ket{z^+}=\ket{0},\ket{z^-}=\ket{1},
\ket{x^\pm}=\frac{1}{\sqrt{2}}(\ket{0}\pm \ket{1})$ and $
\ket{y^\pm}=\frac{1}{\sqrt{2}}(\ket{0}\pm i \ket{1})$,
we find 
\begin{eqnarray}
({\proj{\phi}})^{T_A}
&=& \alpha^2\proj{z^+ z^+}+\beta^2\proj{z^- z^-} \nonumber\\ 
& & +\alpha\beta(\proj{x^+ x^+}+\proj{x^- x^-}     \nonumber\\ & & -\proj{y^+
y^-}-\proj{y^- y^+})\ .  
\label{settings}   
\end{eqnarray} 
Note that this
decomposition contains 6 terms, but only three 
correlated devices' settings.
Alice and Bob have to classically correlate their measurements in their
respective $x,y$- and $z$-directions as indicated in (\ref{settings}),
and add the resulting expectation values with the according positive or
negative weight in order to determine $Tr(W\varrho)$. 
Let us not in passing  that in general an entanglement witness 
does not provide an entanglement measure. However,
for  fixed $ab$
and $d=0$ the probability $p$ in Eq. (\ref{rho})
can be found from
the measurement
outcome via the relation $p=(1-4 Tr(W\varrho))/(1+4ab)$.

The pseudo-mixture (\ref{settings}) provides an ONS, since  it is
impossible to decompose $({\proj{\phi}})^{T_A}$ with less than three
devices' settings. Let us assume the contrary, {\em i.e.}
\bea
({\proj{\phi}})^{T_A}&=&
\sum_{i,j=1}^2U_{ij}\proj{u^A_{i}}\otimes\proj{u^B_{j}}+\nonumber \\
&&\sum_{i,j=1}^2V_{ij}\proj{v^A_{i}}\otimes\proj{v^B_{j}}\ ,
\label{2x2decomposition}
\eea 
where $\braket{u_i^A}{u_l^A}=\delta_{il}=\braket{v_i^A}{v_l^A}$ and the 
same holds for $B$. We  expand
$({\proj{\phi}})^{T_A}=\sum_{i,j=0}^{3} \lambda_{ij} \; 
\sigma_i \otimes \sigma_j$, where we denote $\sigma_0=\eins$, and 
\begin{equation}
(\lambda_{ij})=
\left( 
\begin{array}{cccc}
\frac{1}{4}       &0            &0            & \frac{\alpha^2-\beta^2}{4}\\
0                 &\frac{\alpha\beta}{2} &0            & 0                \\
0                 &0            &\frac{\alpha\beta}{2} & 0                \\
\frac{\alpha^2-\beta^2}{4} &0            &0            & \frac{1}{4} 
\end{array}
\right) \ .
\end{equation}
Note that the $3\times3$ submatrix in the right bottom corner is of rank
three.  Now we write any projector in the rhs of (\ref{2x2decomposition})
with  Bloch vectors: the projector 
$\proj{u^A_{1}}=\sum_{i=0}^3 s^A_i \sigma_i$ is 
represented by the vector $\vec{s}^A=\half(1,s^A_1,s^A_2,s^A_3),$ 
and $\proj{u^A_{2}}$ by $\vec{s}_\perp^A=\half(1,-s^A_1,-s^A_2,-s^A_3)$.
Expanding the first sum on the rhs of 
(\ref{2x2decomposition}) in the ($\sigma_i \otimes \sigma_j$) basis 
leads to a $3\times 3$ submatrix in the right bottom corner which is
proportional to
$(U_{11}-U_{12}-U_{21}+U_{22}) (s^A_1,s^A_2,s^A_3)^T
(s^B_1,s^B_2,s^B_3)$, and thus of rank one.
The corresponding matrix from the second sum on the rhs 
of (\ref{2x2decomposition}) is also of rank one, and we arrive at a
contradiction: no matrix of rank three can be written as a sum of two
matrices of rank one. \hfill $\Box$

We now emphasize the power of  witness operators as a tool 
for the detection of entanglement by discussing the noise in Eq. (\ref{rho}) 
in some more detail. When $d\neq 0$, the state $\varrho$  lies within a 
ball  $B_{p,d}$
with radius $(1-p)d$. If $p$ is such that this ball is 
either included in the set of separable states, or in 
the set of entangled states, then the given $W$ is optimal
and the sign of $Tr(W\varrho)$ provides a signature
of entanglement versus separability. If, however, 
the ball lies across the boundary between those two sets, 
errors may occur. Different questions can then be addressed:
first, one may  want to be sure that a given state
is separable. 
For the case  $a=b=1/\sqrt{2}$ (which we assume here and in the following)
we can estimate a lower bound  $\tau$ such that  
if $Tr(W\varrho)\geq\tau$ then $\varrho(p,d)$ is necessarily separable.
This bound, which
depends on $d$, is
given by: 
\be
\tau(d)=\frac{1}{4}-d^2-\sqrt{(\frac{1}{12}-d^2)(\frac{3}{4}-d^2)}.
\label{lowerbound} 
\ee
For any $\tau'$ with  $0\le \tau'< \tau$, there exists an entangled state
$\varrho(p,d)$ with $Tr(W\varrho)=\tau'$. To derive Eq. (\ref{lowerbound}), 
one uses the fact  that there is a ball $\mathcal{B}$ of separable states
of maximal radius $1/\sqrt{12}$ around $\eins/4$.
Since we do not know $p$ we can only say that 
$\varrho \in \bigcup_{p\in [0,1]}B_{p,d},$ which is a kind of a convex
cone originating in $\proj{\psi}$ and terminating in the ball $B_{0,d}$ 
of radius $d$ around the totally mixed state.
The bigger $Tr(W\varrho)$, the closer to the ball 
$B_{0,d}\subset \mathcal{B}$ is $\varrho.$
Obviously, if $Tr(W\varrho)$ is big enough we have $\varrho \in \mathcal{B}.$
In this manner one can determine the value of $\tau.$
\hfill $\Box$

\begin{figure}[h]
\centerline{\psfig{figure=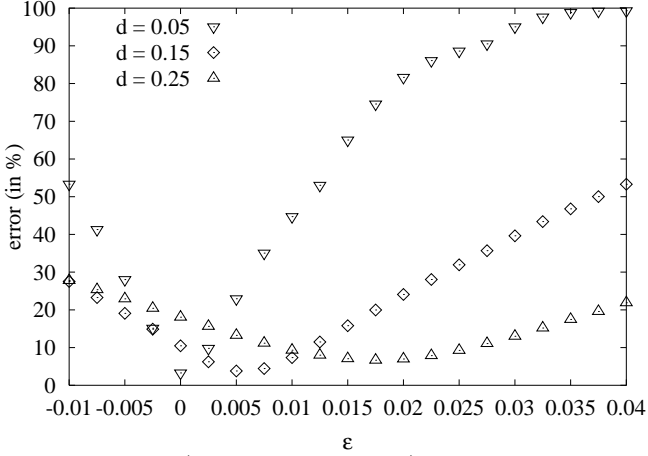,width=0.5\textwidth}}
\caption{\small Error   (maximised over all $p$)
as a function of $\varepsilon$, for three different values
of $d$. }
\label{fig1}

\end{figure}

Second, one may be interested in minimizing the 
probability of making an error, either by  
mistaking a separable state for an entangled one, or vice versa. 
If we assign $Tr(W\varrho)>0 \Leftrightarrow \varrho$ separable,
it turns out that -- depending on the value of $d$ -- in order to
minimize this error, it is more favorable to use 
$W_{\varepsilon}:=W-\varepsilon \eins$.
This operator is not a ``witness" in the original sense, because 
it yields negative expectation values for some separable states.
Note that it requires the same measurement settings as $W.$
To estimate the error inherent in this detection scheme, we have 
used the method from \cite{volumes2} to randomly 
generate a sample of 50000  density matrices of the 
form (\ref{rho}), and then checked their separability using 
both the partial transposition criterion and applying $W_{\varepsilon}$. 
 The percentage of  errors when using $W_{\varepsilon}$ is plotted in 
Fig. \ref{fig1}.  For large $d$ the operators $W_{\varepsilon}$ are  in fact 
less erroneous in detecting entanglement. 
Further numerical analysis suggests that the  optimal $\varepsilon$  
increases quadratically with $d$.


Generalizing the above results to higher dimensions 
for states with non-positive partial 
transpose (NPT states) is possible, although not straightforward. 
In an $N\times M$ dimensional Hilbert space with $N\le M$, 
we first have  to identify the vector $|\phi\rangle$ that corresponds to the 
minimal (negative) eigenvalue of $\varrho^{T_A}$. Without loosing 
generality we may assume that it has  maximal Schmidt rank, and is given by  
$|\phi\rangle=\sum_{i=0}^{N-1}\alpha_i|ii\rangle$. Obviously, the ONP 
corresponding to 
$(|\phi\rangle\langle\phi|)^{T_A}$ must contain at least $N^2$ terms since 
the rank of $(|\phi\rangle\langle\phi|)^{T_A}$ is $N^2$. For $N=2$, {\em i.e.} 
for $2\times M$ dimensional systems the results obtained for $M=2$  
are also valid for $M>2$, since the maximal Schmidt rank in such spaces is 2.
This implies that the ONP must contain 5 terms, 
whereas the ONS can be realized with 3 settings.
For $N\times M$ systems with $N\ge 3,\ N\le M$ 
we can easily construct a pseudo-mixture with $2N^2-N$ terms
using the same method as in the case of $2\times 2$ systems.
This gives  the upper bound for the number of terms in ONP, 
and corresponds to an upper bound for ONS of $2N-1$ $(2N)$ for
even (odd) $N$. By generalizing the method used to demonstrate that 
two settings are not enough in $2\times 2$ to the $N\times M$ 
case one can prove that any ONS must contain at least 
$N+1$ settings. It is not clear, however, 
whether this bound can be reached in general. 

In higher dimensions there also exist entangled states with 
positive partial transpose, namely bound entangled 
states~\cite{pawel67,boundprl}. For this type of states no general 
operational entanglement criterion is known, and thus even the full 
knowledge of the density matrix may not suffice to decide whether
the density matrix is entangled or not. 
There exists, however, an important class of
bound entangled states, the so-called ``edge" states, for which  optimal 
witness operators can be constructed explicitly.  
A state $\delta$ is called an edge state iff it cannot be represented as
$\delta=q\delta'+(1-q)\sigma$, where  $\sigma$ is a separable 
state, $\delta'$ is a state with positive partial transpose,
and $0\leq q<1$. Edge states 
are, in a certain sense, the bound entangled analogues 
of pure entangled states. 
In the situation where an experiment is aimed at the generation of 
an edge state,
our method of local decomposition of a witness 
provides  a genuine
experimental test.  

Let us illustrate this with the example of 
unextendible product basis (UPB) states~\cite{Bennet99} in a
$3\times 3$ dimensional  space. 
The states
\begin{eqnarray}
\ket{\psi_{0}}&=&\frac{1}{\sqrt{2}}\ket{0}(\ket{0}-\ket{1}), \hspace{0.5cm} 
\ket{\psi_{2}} = \frac{1}{\sqrt{2}}\ket{2}(\ket{1}-\ket{2}),\nonumber\\ 
\ket{\psi_{1}}&=&\frac{1}{\sqrt{2}}(\ket{0}-\ket{1})\ket{2}, \hspace{0.5cm} 
\ket{\psi_{3}} = \frac{1}{\sqrt{2}}(\ket{1}-\ket{2})\ket{0},\nonumber\\ 
\ket{\psi_{4}}&=&\frac{1}{3}(\ket{0}+\ket{1}+\ket{2}) 
(\ket{0}+\ket{1}+\ket{2})
\end{eqnarray} 
form a UPB, {\em i.e.} they are orthogonal to each other and 
there exists no product vector which is orthogonal to all of them.
The state
$
{\rho_{BE}}=\frac{1}{4}(\eins-\sum_{i=0}^{4}\proj{\psi_{i}}) 
$
constructed from this UPB is an entangled state with positive 
partial transpose.
The generic form of an entanglement witness 
for such a state is\cite{terwit,opti}
\begin{eqnarray}
W&=&(P+Q^{T_A})/2-\epsilon\eins\ ,\label{ndwit} \\ 
\epsilon&=& \inf_{\ket{e,f}}\bra{e,f}P+Q^{T_A}\ket{e,f}/2 \ ,
\nonumber
\end{eqnarray}
where 
$P$ and $Q$ denote the projectors onto the kernel of $\rho_{BE}$ 
and the kernel of $\rho_{BE}^{T_A}$, respectively.
For the given UPB state we have
$
P=Q^{T_A}=\sum_{i=0}^{4}\proj{\psi_{i}}$.
The main problem for the construction of $W$ is to find $\epsilon$.  
An analytical bound obtained by Terhal\cite{terwit} gives 
$\epsilon \ge 0.0013$.
Numerical analysis leads however to the  much bigger 
value $\epsilon \simeq 0.0284$.
Once $\epsilon$ is found, the decomposition of $W$ is straightforward:
the explicit form of the witness  fixes five elementary measurements, 
and the identity can be decomposed into nine orthogonal projectors onto 
product vectors 
such that four of them coincide with four of the UPB states. This follows 
from the fact that the vectors $|\psi_{0}\rangle$, $|\psi_{1}\rangle$, 
$|\psi_{2}\rangle$, $|\psi_{3}\rangle$ can be extended to an orthonormal 
basis by defining
\begin{eqnarray}
\ket{\bar\psi_{4}}&=&\frac{1}{\sqrt{2}}\ket{0}(\ket{0}+\ket{1}), 
\hspace{0.5cm} \ket{\bar\psi_{5}} = \frac{1}{\sqrt{2}}\ket{2}(\ket{1}+\ket{2}),
\nonumber\\ \ket{\bar\psi_{6}}&=&\frac{1}{\sqrt{2}}(\ket{0}+\ket{1})\ket{2}, 
\hspace{0.5cm} \ket{\bar\psi_{7}} = 
\frac{1}{\sqrt{2}}(\ket{1}+\ket{2})\ket{0},
\nonumber\\ \ket{\bar\psi_{8}}&=&\ket{1}\ket{1}.
\end{eqnarray}
Altogether we are left with a pseudo-mixture that contains 10 projectors.
Denoting by $B_1=\{\ket{0},\ket{1},\ket{2}\}$, $B_2=\{(\ket{0}-\ket{1})/
\sqrt 2,
\ket{2},(\ket{0}+\ket{1})/\sqrt 2\}$,  $B_3=\{(\ket{1}-\ket{2})/\sqrt 2,
\ket{0},(\ket{1}+\ket{2}/\sqrt 2)\}$, and $B_4=\{(\ket{0}-\ket{1})/\sqrt 2,
(\ket{0}+\ket{1}+\ket{2})/\sqrt{3}, (\ket{0}+\ket{1}-2\ket{2})/2\}$, 
we easily see that measurement of $W$ for this decomposition 
requires 6 correlated settings for Alice and Bob: $B_1B_2$, $B_2B_1$, 
$B_1B_3$, $B_3B_1$, $B_4B_4$, and $B_1B_1$. This result
implies that the ONS must be $\le 6$. By subtracting in Eq. (\ref{ndwit})
some positive operator $I$ of full rank instead of the identity $\eins$, 
one can reduce the number of projectors in the decomposition of $W$
 to 9, and this gives an ONP, since the number 
of terms in any ONP must be larger or equal than the rank of the witness, 
which is equal to 9. The idea is to form $I$ as a convex sum of 
projectors onto $|\psi_i\rangle_{i=0,\ldots,4}$ and onto
 4 other product vectors 
that are obviously  not orthogonal to 
$|\psi_i\rangle_{i=0,\ldots,4}$, but such that the set  
of the 9 vectors
forms a basis. 
Here the bound for $\epsilon$ has to be adapted to the choice of $I$,
such that positivity of $W$ on all separable states is guaranteed,
i.e.
\begin{equation}
 \epsilon'=\inf_{|e,f \rangle}
   \frac{\langle e,f|(P+Q^{T_{A}})/2 |e,f\rangle}
   {\langle e,f| I |e,f\rangle}.
\end{equation}

If we choose as additional vectors
$|\bar\psi_i\rangle_{i=4,\ldots,7}$
the decomposition contains 9 projectors in only 5 settings.
Numerical analysis leads to $\epsilon' \simeq 0.0311$
for this choice of $I$.
Note that when the  bound entangled state is 
affected by white noise, namely
$
\rho_{p}=p\cdot\rho_{BE}+(1-p)\eins/9 ,
$
the witness given above is still suitable for the detection of
entanglement, provided
$Tr(W\rho_{p})<0$. 
For the witness in Eq. (\ref{ndwit})
this is the case  when $ p>(1-9\epsilon/5)$.
Let us mention that for experimental
purposes it is not necessary to decompose 
the identity in Eq. (\ref{ndwit}) since this term will only add
a constant ($-\epsilon$) to the outcome of the measurement 
of the ``prewitness'' $\bar{W}=(P+Q^{T_A})/2.$ This implies  that indeed
one only needs to find the corresponding  
ONP and ONS for $\bar{W}$. In this way one obtains 
an ONP with only 5 terms, and that the ONS 
has to be less or equal to 5.

In summary, we have introduced optimal
decompositions  of witness operators into {\em local} 
projectors for the
detection of  entanglement. This method 
can be used with present experimental techniques. 
It is a very powerful method to detect entanglement in the cases
where one has a certain knowledge about the state that one wants
to create, e.g. when the aim is to produce a 
specific pure entangled state,
but this state is corrupted by noise. At the
present stage of quantum information processing, several
experiments strive at creating such entangled states, and thus
it is important to show that the produced state is indeed entangled. 
The more knowledge about the state is given, the less knowledge about
the underlying noise is necessary for unambiguous classification of
the state. For the situation in which little knowledge about the created
state exists it will be favourable and maybe necessary to utilize 
more than one witness operator. 
A detailed analysis of the trade-off between the initial knowledge 
of the state and the witnesses needed is left for further research.

After  submission  we became aware of a recent preprint
by A. Pittenger and M. Rubin \cite{piru} where the ideas of this paper
have been further developed. In particular it is shown there 
that if $N$ is prime, a projector onto a maximally entangled 
state with full Schmidt rank can be measured with $N+1$ local 
measurements, so our bound can be reached for this case.

We wish to thank I. Cirac, S. Haroche, S. Huelga, B. Kraus, 
H. Weinfurter, and K. \.Zyczkowski for discussions.
This work has been supported  by  DFG 
(Graduiertenkolleg 282 and
Schwer\-punkt  ``Quanteninformationsverarbeitung"), 
the ESF-Programme PESC,  and the EU IST-Programme
EQUIP.


\begin{thebibliography}{99}
\bibitem{stern} For detection of spin $S$ particles  a Stern-Gerlach
device can be used, for which a setting corresponds to the choice of the
direction of the magnetic field and detection of
 each of the possible $2S+1$ outputs.
\end{thebibliography}
\end{document}